# Knowledge-Base Practicality for Cybersecurity Research Ethics Evaluation

Robert B. Ramirez[1,a)]  Tomohiko Yano[1]  Masaki Shimaoka[1]
Kenichi Magata[1]

**Abstract:** Research ethics in Information and Communications Technology has seen a resurgence in popularity in recent years. Although a number of general ethics standards have been issued, cyber security specifically has yet to see one. Furthermore, such standards are often abstract, lacking in guidance on specific practices. In this paper we compare peer-reviewed ethical analyses of condemned research papers to analyses derived from a knowledge base (KB) of concrete cyber security research ethics best practices. The KB we employ was compiled in prior work from a large random survey of research papers. We demonstrate preliminary evidence that such a KB can be used to yield comparable or more extensive ethical analyses of published cyber security research than expert application of standards like the Menlo Report. We extend the ethical analyses of the reviewed manuscripts, and calculate measures of the efficiency with which the expert versus KB methods yield ethical insights.[a]

## 1. Introduction

Research ethics in Information and Communications Technology (ICT) has seen a resurgence in popularity in recent years, spurred in part by Artificial Intelligence (AI) [6,12]. Although a number of general standards have been issued in the past decade, there are currently no easily usable, granular, or comprehensive benchmarks for evaluating the ethics of cyber security research projects. There is also currently no method for evaluating security research ethics in a truly systematic or reproducible manner [2,5,20,23,32,36].

Existing frameworks for ethics in ICT research are abstract in that they either focus on the ethical assessment process as a whole, give only general advice, or do not focus on security. In reality, researchers have to deal with concrete ethical dilemmas on a variety of topics, as evidenced by the prevalence of 'ethical issues' sections in research papers [16].

As a result, despite ethical analyses being demanded by many top conferences [14], traditionally, committees or Internal Review Boards (IRBs) composed of experts have been necessary to comprehensively review the ethics of papers, which has been seen as a domain requiring significant expertise, often from senior members of the research community [1,8]. However, many IRBs lack experience specifically evaluating ICT research, particularly in a cyber security context [19,36,45]. This frequently leads to ICT research being exempted from IRB review [25,31,40]. Thus, researchers need to be able to evaluate their own research. This poses a barrier for those with less experience with ethics reviews.

In this paper we perform a best-case scenario test of the ability of security researchers without ethics committee experience to evaluate research ethics. We employ a knowledge base (KB) of concrete cyber security research ethics best practices, compiled in [30,41] from a large random survey of research papers. We contribute 1) evidence that such a KB can be used by non-experts to yield ethical analyses of published cyber security research comparable to those current benchmarks yield in the hands of experts, and 2) a novel approach for comparing these two analysis methods.

We assess two separate ethically-condemned research papers and corresponding published ethical analyses written by experts about them, and show that the principles collected in the knowledge base yield, depending on the calculation method, between 3 and 12 times as extensive an analysis as those given by the experts, after accounting for redundant and irrelevant information. We define and measure the coverage and the efficiency of both the expert reports and the KB-based analyses by systematically extracting and organizing the claims made in the expert reports and those made with the KB, to make them comparable. The KB analyses encompassed the vast majority of the observations found in the expert reports, and further yielded a number of novel insights expanding on the ethics of the original papers.

[1]  SECOM CO., LTD. Intelligent Systems Laboratory, Mitaka, Tokyo, Japan
[a)] ro-ramiresu@secom.co.jp

## 2. Background

### 2.1 ICT Ethics Guidelines and Related Work

There are a number of existing ICT guidelines relevant to cyber security research ethics. Here we highlight some of the more well known ones and their relation to our work.

The Menlo Report was released in mid-2012 as 'guidance for ICT researchers' in 'the context of ... information security research' [20]. It was inspired by the Belmont Report developed in the 1970s for medical ethics. When formal ethical discussions are included in cyber security research papers, the Menlo Report or its principles are sometimes referenced [22]. It includes detailed high-level guidance on the types of considerations necessary when conducting ethical research, including stakeholder analysis, respect for persons, beneficence, justice, and the public interest.

The Menlo Report gives a few example applications of its principles to cyber security research, with hypothetical examples that involve phishing, vulnerability disclosure, and handling sensitive information; but it does not describe the details of the ethics of those actions themselves. While the Menlo Report and its *Companion* [19] provide a methodology for ethics, they do not commit to recommending concrete best practices for specific activities. *The Companion* does, however, include a synthetic case study that goes into detail about how to apply its guidelines to a few specific hypothetical research actions, but it is limited in scope, organized as prose, and lacks specific references to research.

A tool called 'CREDS' was at one point in development by a disjoint set of researchers affiliated with the authors of the Menlo Report and the U.S. Department of Homeland Security [11]. The stated goal of this tool, first proposed in 2015, seems similar to [30, 41] (the subject of this paper) in that it seeks to somehow analyze best practices, as well as laws, to create an online ethics tool [33].

### 2.2 Cyber Security Ethics Knowledge Base

In [30, 41], the authors extracted descriptions of ethical practices from 101 relevant papers out of a collection of 943 published in the top conferences in cyber security between 2013-2017. The findings were compiled into a knowledge base (KB) using a decision tree structure (Figure 1) [3, 41].

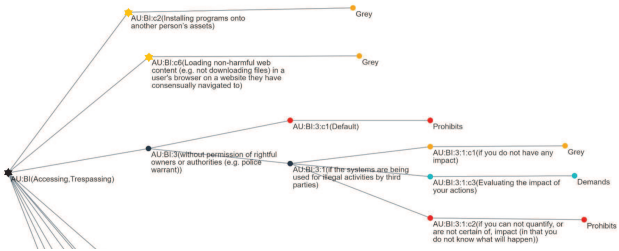

Fig. 1: Decision Tree Sample

#### 2.2.1 Structure

Each branch of the tree terminates with leaves that, along with their branches, specify the ethics of actions ($\mathbb{A}$) a researcher might consider taking. A given path from root to leaf is meant to be human-readable as a full statement of actions. There are 5 types of leaves: "Permitted", "Prohibited", "Demanded", "Gray," and "Recommended," based on traditional Deontic Logic [39]:

(1) *Permitted* means performing $\mathbb{A}$ is not in itself unethical
(2) *Prohibited* means performing $\mathbb{A}$ is in itself unethical
(3) *Demanded* means not performing $\mathbb{A}$ is itself unethical
(4) *Gray* means $\mathbb{A}$ could be either *Permitted* or *Prohibited*
(5) *Recommended* indicates *Permitted* or *Demanded*.

*Gray* and *Recommended* are "*TBD*" placeholders indicating a lack of consensus on the ethics of $\mathbb{A}$.

## 3. Ethically-Condemned Manuscripts

In this paper we use [30, 41]'s KB to analyze the ethics of specific research actions described in research papers that have been condemned as unethical by the community, and compare our analyses to the corresponding published expert ones, which used the Menlo Report and the Association of Internet Researchers (AoIR) Recommendations [36]. Here we give a description of the papers and their controversies, based in part on our own analyses (Section 6).

### 3.1 Internet Census 2012

Internet Census 2012 (Carna Botnet) [17] is a 2012 non-peer-reviewed paper by an anonymous hacker about an illegally-conducted survey of insecure machines [15, 34, 44].

The Carna Botnet creator built a port scanner to scan the majority of the Internet. Its efficiency was enabled by piggybacking off of insecure devices to propagate their malware-like program, exponentially increasing the number of nodes performing scans. The paper was published before other efficient and comparatively more ethical solutions like Zmap [27], so it may have been considered novel at the time.

Among the ethical malpractices included in Internet Census 2012's research are not only breaking into devices, but also releasing all the data from the scans. Because some have considered Carna a "nice" botnet, there has been discussion about whether it is ethical to use its data in mainstream research [4, 38, 43]. The Internet Census 2012 author made some effort to reduce the burden imposed on devices by limiting the number of simultaneous connections the scanner can make, avoiding probing further into networks beyond insecure routers, and even removing malware found on devices if it interfered with Carna's scans. However, the laundry list of illegal and unethical practices in deploying Carna made the paper an immediate target for condemnation.

### 3.2 Encore

Encore: Lightweight Measurement of Web Censorship with Cross-Origin Requests [21] details an experiment in which the authors released a program for web hosts that uses cross-origin requests to fetch content from separate websites that are allegedly censored in certain countries, in order to confirm what is censored to whom. Amassing data from an array of vantage points in this way enables the authors to perform longitudinal measurements of censorship.

The controversy in the paper lies mainly in the fact that unsuspecting users navigating to otherwise safe websites could be served censored content, putting them at risk with governments. Furthermore, visitors' information, such as their IP address (considered personal identifiable information under the GDPR [13, 24]), is sent directly to the authors of Encore, without obtaining users' informed consent.

The official version of the paper on SIGCOMM's website includes a statement from the program committee (PC) stating that while it appreciates the technical contributions of the paper, there are a number of ethical concerns with the experiments the authors conducted that lead the PC to be unable to endorse it, and that the controversy surrounding the paper arose in large part because of the lack of ethics standards for this type of research. The statement continues that because the authors engaged with their IRB and yet the IRB did not flag the research as unethical, the PC elected to publish the paper as a case study for the community.

## 4. Proposal

To enable non-experts to perform ethics analyses that more efficient and comprehensive than experts', we propose that it ought to be feasible to compile a knowledge base with sufficient mention of ethical issues (coverage) for such a purpose by sampling from research papers and ethics standards. In this paper we use a KB compiled from peer-reviewed papers [30, 41]. If the KB has a concrete level of granularity and much greater coverage (as compared to current abstract standards) based on issues researchers actually encounter, use of such a KB should yield more comprehensive and efficient ethical analysis results than expert reviews using current ethics standards. This is because the KB is based on similar cases to the research under evaluation.

Current standards require significant extrapolation to draw conclusions from, and despite their existence there are still many ethical questions in the security research community, indicating a deficiency in guidance [9, 10].

Presumably, committees formed from people with significant experience reviewing research papers and applying standards to different cases are the gold standard for ethical review [1, 8]. These experts likely gained much of their expertise by reviewing research papers and proposals. Therefore, we posit that a good measure of the quality of the KB is how well it performs against such an expert armed with accepted standards like the Menlo Report. To account for the influence of experience, a member of our team with no experience serving on review committees performed the analysis that used the KB. Our comparison includes evaluations of both comprehensiveness and information added, which accounts for noise and redundancy.

## 5. Experiment

In this section we outline the experiment we performed to compare the coverage and efficiency of the KB versus expert review. For this experiment our notion of efficiency includes a large percentage of reported observations actually being related to ethics (low noise), and a low percentage of repeated or rephrased observations (low redundancy). In this paper we use the terms *item* and *observation* mostly interchangeably. An *item* is a specific identifiable point we use for quantifying analyses; an *observation* refers more generally to the concept of an ethic issue identified in an item.

### 5.1 Experiment Procedure

Since an ethical analysis with more comprehensive coverage of ethics-related observations can be considered better, we performed an experiment consisting of three parts in order to determine how the coverage (after accounting for noise and redundancy) of analyses yielded by the KB compares to that obtained when using current standards (Figure 2). One can imagine that, at a minimum, obtaining the same result as an expert analysis would be a successful result, given that the KB analysis is performed by an ethics amateur.

To assess coverage we compare *the number of ethical observations* yielded by the KB to those made by experts in published ethical critiques for the same research projects.

To eliminate noise in the analyses, a blind majority vote between three researchers is used to classify out-of-scope items. We report the coverage from each approach before and after accounting for redundant items by determining whether an items' observations are covered by other ones.

We reason that if the noise- and redundancy-adjusted coverage obtained from using the KB matches or exceeds that of the traditional expert reviews for all case studies, then the KB approach is an effective ethical analysis method.

#### 5.1.1 Paper Selection

In 2018 we performed a literature review to find applications of ICT or security ethics standards to case studies. In our literature review we were unable to find case studies of any substantial scale that used ethics standards besides the Menlo Report. As a result, we limited our search to critiques published after 2012, which could have used the Menlo Report in their analyses. This review turned up only two critiques: ones of the papers in Section 3. Both used the Menlo Report as the basis for their analysis. The Encore Report also drew on the AoIR's ethical recommendations.

The Internet Census 2012 critique was written by an author of the Menlo Report and is peer-reviewed. The Census itself was published only shortly after the Menlo Report, so the author may have been unable to refer to it, however.

The Encore critique is a federally-funded paper written by the person who first pointed out Encore's ethical issues to ACM SIGCOMM, giving it some form of peer-acceptance. It has been cited more than the Internet Census 2012 ethical critique. In this experiment we analyze both the original

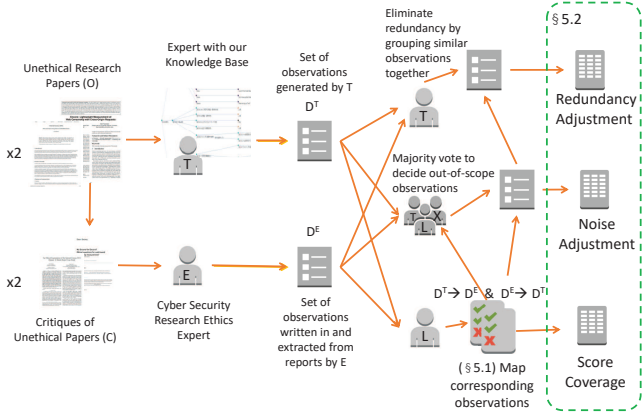

Fig. 2: Comparing Analyses Using a Knowledge Base Versus Existing Standards

research papers, **O**, and their published ethical critiques, **C**.

The ACM Code of Ethics (CoE) also has on its website very short hypothetical research case studies [7]. We initially considered using these as well, but because of their artificial nature and unrealistic length, we determined that it would be impossible to perform a fair comparison between the KB and the ACM CoE through them. This is not to say that the ACM case studies are poor examples, only that they are not suited for our comparison.

#### 5.1.2 Data Extraction

To assess coverage we compare the number of ethical observations made by **C** and the KB with respect to **O**. There are three main roles members of our team play in the comparison. The first role is a labeler, **L**, who compares the KB approach to the "manual" expert approach. **L** compares the datasets $D^E$ and $D^T$, which are organized sets of observations extracted by analyzing **C** and **O**, respectively.

$D^T$ is the dataset, in spreadsheet format, of ethical observations regarding the ethics of the original papers **O** (one spreadsheet of observations per paper). $D^T$ is created by using the KB to assess the papers without consulting **C**. The creator of $D^T$, fulfilling the second main role, is called **T** (Figure 2), named for the Decision Support (T)ool. In order to ensure a best-case comparison, **T** is someone who has significant familiarity with the KB.

In order to conduct a fair comparison between current standards and [30,41]'s ethical best-practices gathering *methodology* (Section 2.2), for this analysis we excluded the items in the KB that were solely derived from ethics standards, such as the Menlo Report and ACM CoE, and used only those items derived from surveying the literature.

$D^E$ is the set of observations extracted from the ethical critiques **C**. The creator of this dataset, called **E**, for (E)xpert, has significant experience on security ethics and review committees. **E** creates each $D^E$ spreadsheet by reading **C** and recording the authors' noted observations in a spreadsheet format that resembles $D^T$. The role of **E**'s is merely that of a transcriber of the results of the standards-based methodology already applied by the authors of **C**, who have a degree of expertise in ICT or cyber security ethics. Therefore **E** is instructed to refer to **O** only for clarification.

#### 5.1.3 Mapping

A third party **L** labels the unique and shared observations from $D^E$ with respect to $D^T$, then repeats going from $D^T$ to $D^E$, for each of the two papers. In each case, the labeling is said to go from the "primary" set to the "secondary set," whose items are referenced in the labels. **L** does not read **C**, and only reads the introductions from **O**, for context. Our preliminary tests showed this basic background knowledge to be necessary for comparing items.

This two-sided mapping is done for two reasons. First, tractability, because each set of analysis results contained on the order of 30 to 90 results, requiring a systematic approach for consistency in labeling. This simplifies the scoring of the final results. Second, to mitigate bias arising from considering matches from the point of view of only one side.

- $\nvdash$: Used if no other labels apply. The item is in-scope, and no items in the secondary set correspond to it.
- $+\alpha$: There is some similarity, shared implied meaning, or overlap between the items in question. Specifically, the secondary item is included in the primary one, which adds some value. **L** must be able to articulate the remaining *difference*, though, or the label becomes $\emptyset$.
- $\emptyset$: The intended meaning of the primary item seems similar to the extent that the difference between the secondary item cannot be easily articulated. This label should be used even if the primary item is "$-\alpha$" with respect to the secondary item - that is, even if it is entirely included in the (inferred) intended meaning of the second item. This is because comparisons are done one-sidedly, such that only the *primary set* matters for this label. The secondary set's contributions are enumerated when sets are reversed.
- $\notin \mathbb{S}$ *(determined by a vote of three)*: The item did not contain a description of any of: something researchers ought to do or take into consideration, an explicit statement that something is unethical or obligatory, or a risk of a negative externality to a party. For fairness, **L** was instructed to label items that appear $+\alpha$ or $\emptyset$ as in-scope, regardless. For brevity we use the set notation $\in \mathbb{S}$ to mean in-scope and $\notin \mathbb{S}$ for out-of-scope.

### 5.2 Scoring
#### 5.2.1 Coverage

Each of $D^E$ and $D^T$ is evaluated based on its number of unique ethics observations between the two datasets, by assigning points to it. The labeling system (above) is divided into labels that award points, and labels that do not.

$\mathbb{S}^+$ (Awards Points) items include $\nvdash$ and $+\alpha$ and contain some unique value-added ethical observation by the primary set. Because a brand new item ($\nvdash$) is more unique than a slightly expanded item ($+\alpha$), one may wish to weight these two types of items differently. Within these two types, all contributions are awarded 1 point equally. To avoid bias against items with multiple observations, during the creation of $D^E$ and $D^T$, **E** and **T** made efforts to ensure that all items are maximally subdivided, to eliminate these cases.

$\mathbb{S}^0$ (Does Not Award Points) labels include $\notin \mathbb{S}$ and $\emptyset$, and indicate no *unique* value added by the item, and do not result in points for the approach the primary set represents.

#### 5.2.2 Noise

To eliminate "noise" in the analyses, we had two additional researchers besides **L** independently label out-of-scope nodes, and we then used the majority decision between those three researchers to classify items as $\in \mathbb{S}$ or $\notin \mathbb{S}$.

Having this vote is necessary since coverage results depend heavily on whether an item is labeled $\in \mathbb{S}$. Since the KB should theoretically have no out-of-scope items, for a given $\notin \mathbb{S}$ item in $D^E$, if there is no corresponding ($\notin \mathbb{S}$) item in $D^T$, the primary other label competing for that item would be $\nvDash$, so such items are highly likely to subsequently be categorized as $\nvDash$ by **L**. This means $D^E$ could gain an advantage if it presents a lot of $\notin \mathbb{S}$ items hoping to get some past **L**.

$\notin \mathbb{S}$ items were especially prevalent in the Encore critique. This was a result of it being structured to include a general overview of the Encore program in the first half of the paper.

#### 5.2.3 Redundancy

Throughout the experiment it became clear that a substantial number of items in $D^E$, and a few items in $D^T$, were completely or significantly redundant, resulting in a disconnect between the granularity of claims in the two sets. To ensure we measure the amount of *novel* information each ethical analysis approach contributes, we also compared $D^E$ and $D^T$ after accounting for redundant observations.

To eliminate this within-set redundancy and present all items at similar level of granularity, a researcher (**T** in this experiment), separate from the primary labeler **L**, grouped together similar items within each of the four sets (two from $D^E$, two from $D^T$). This was done after the initial labeling.

Although in many cases redundant items within a set happened to have the same labels, the following rules were then used to select the final label for a given set of redundant items that had multiple labels:

(1) If an item is a member of multiple redundant groups, the label associated with that item is not a reliable indicator of the individual group's label, since it is unclear which group that label actually applies to.
(2) If one of the items was linked to an extremely large number of items in the secondary set, its label was not used as it could be expected to be too broad.
(3) Connections between items were then emphasized, yielding the priority $(+\alpha \to \emptyset) \to (\notin \mathbb{S} \to \nvDash)$, where items labeled $+\alpha$ and $\emptyset$ had priority because there were items in the secondary set associated with them, unlike $\notin \mathbb{S}$ and $\nvDash$. The highest priority label among a group's items was chosen as the final label.
(4) If there were an equal number of singular items labeled as $\emptyset$ and $\nvDash$, as a compromise $+\alpha$ was selected.

We prioritized $+\alpha$ over $\emptyset$ because if a group contributed $+\alpha$ from one item solely belonging to that group, and another item that did not contribute a $+\alpha$ (i.e. that was only $\emptyset$), then we assume that '$+\alpha$' + '$\emptyset$' = '$+\alpha$' the same way that $1 + 0 = 1$, using the reasoning that '$\mathbb{S}^+$' + '$\mathbb{S}^0$' = '$\mathbb{S}^+$'. By the same logic we treated '$\nvDash$' + '$\emptyset$' = '$\mathbb{S}^+$', but we equated this to $+\alpha$ because, due to the $\emptyset$, there were still clearly some aspects of the item that were the same, and by the definition of $+\alpha$ given in Section 5.1.3, the focus is on whether there is some similarity between the items (as there is with $\emptyset$). As described in Section 5.1.3, $\nvDash$ was only selected as a last resort (regardless of whether **T** or **E** was the primary set). For $\notin \mathbb{S}$ and $\nvDash$, because of the risk mentioned above regarding noise, we prioritized $\notin \mathbb{S}$.

## 6. Results

We report adjusted results to the calculated coverage after accounting for noise and redundancy sequentially. By comparing original vs adjusted coverage we can understand the raw amount of reported observations from each approach as well as how efficient each is at generating observations.

Table 1. Observations from Expert Versus Knowledge Base

|  | Raw Score | | | No $\notin \mathbb{S}$ | | | No Redundancy | | |
|---|---|---|---|---|---|---|---|---|---|
|  | E | T | T/E | E | T | T/E | E | T | T/E |
| *Census* | | | | | | | | | |
| $\in \mathbb{S}$ | 26 | 36 | *1.4* | 23 | 36 | *1.6* | 13 | 25 | *1.9* |
| $\nvDash$ | **10** | **19** | | **8** | 19 | | **4** | **12** | |
| $+\alpha$ | **9** | **16** | | **8** | 16 | | **5** | **12** | |
| $\emptyset$ | 7 | 1 | | 7 | 1 | | 4 | 1 | |
| $\notin \mathbb{S}$ | 15 | 0 | *0* | 18 | 0 | *0* | 18 | 0 | *0* |
| G[1] | **41** | **36** | *.88* | 41 | 36 | *.88* | **31** | **25** | *.81* |
| $\mathbb{S}^+$[2] | **19** | **35** | *1.8* | **16** | 35 | *2.2* | **9** | **24** | *2.7* |
| %N[3] | .46 | .97 | *2.1* | .39 | .97 | *2.5* | .29 | .96 | *3.3* |
| *Encore* | | | | | | | | | |
| $\in \mathbb{S}$ | 41 | 34 | *.83* | 30 | 34 | *1.1* | 13 | 25 | *1.9* |
| $\nvDash$ | 7 | **20** | | **1** | 20 | | **0** | **12** | |
| $+\alpha$ | 13 | 5 | | **12** | 5 | | **4** | **7** | |
| $\emptyset$ | 21 | 9 | | 17 | 9 | | 9 | 6 | |
| $\notin \mathbb{S}$ | 40 | 0 | *0* | 51 | 0 | *0* | 51 | 0 | *0* |
| G[1] | **81** | **34** | *.42* | 81 | 34 | *.42* | **64** | **25** | *.39* |
| $\mathbb{S}^+$[2] | **20** | **25** | *1.3* | **13** | 25 | *1.9* | **4** | **19** | *4.8* |
| %N[3] | .25 | .74 | *3.0* | .16 | .74 | *4.6* | .06 | .76 | *12* |

[1] $T := \in \mathbb{S} + \notin \mathbb{S}$;  [2] $\mathbb{S}^+ := \nvDash + \alpha$;  [3] $\%N(\%NEW) := T/\mathbb{S}^+$

Table **1** shows the detailed results of labeling the ethics observations, for both the Internet Census (top) and Encore (bottom). Results are reported to two significant figures. *Raw Score*, *No $\notin \mathbb{S}$*, and *No Redundancy* indicate the raw number of items found by **E** and **T** (calculated from **L**'s labeling of $D^E$ and $D^T$) for unadjusted coverage, the adjusted numbers after accounting for noise, and the adjusted numbers after further accounting for redundancy, respectively. The **T/E** (%) columns show a comparison of the two analysis approaches. For numbers of interest, they give the ratio **T:E** as a percentage of the two columns immediately to the left of each of the respective **T/E** (%) columns in table **1**.

The $\in \mathbb{S}$ rows show the sum of all the $\nvDash$, $+\alpha$, and $\emptyset$ items for a given analysis. *(G)* shows the total number of items using each of the **E** and **T** approaches. They are the sum of their respective $\notin \mathbb{S}$ and $\in \mathbb{S}$ items for each of the two papers (light blue). The value in row *(G)* always remains unchanged until redundancy is accounted for, when similar observations are combined into a single item.

The top light-gray rows show the sums of all $\mathbb{S}^+$ items ($\nvdash$ and $+\alpha$) for each stage of the calculation. These rows give an indication of the total coverage of ethical issues yielded by each approach. $\emptyset$ items are contributed by both approaches and therefore do not award any points. The bottom light gray rows (% NEW) are a measure of the efficiency of each of the approaches. They show, as a percentage, the fraction of items reported by each approach that are unique to that approach, rather than being shared ($\emptyset$) or irrelevant ($\notin \mathbb{S}$).

Bold numbers indicate changes across columns. For example, in *No* $\notin \mathbb{S}$, the number of items from **T** remained unchanged, but **E** saw a slight decrease in their $\in \mathbb{S}$ items. The individual changes to the $\nvdash$ and $+\alpha$ rows are shown in bold, but to avoid cluttering the table, the overall changes are not repeated in $\in \mathbb{S}$ and $\notin \mathbb{S}$ rows.

### 6.1 Summary of Results

Table 1 shows that the knowledge base approach yielded substantially more novel information in the form of ethics observations than the expert critiques based on existing standards, despite yielding fewer observations overall.

Using $D^T$, the Internet Census analysis shows gains ranging from 80% (i.e. a **T/E** value of 180%) to 170% (i.e. a **T/E** value of 270%), and the Encore analysis shows gains ranging from 30% to 380%, depending on whether noise and redundancy are accounted for. Gains are determined by subtracting 100% from the calculated value of the ratio **T/E**.

In order to consider ethical observations as independent events for statistical purposes, we must only consider unique observations, so for that purpose it only makes sense to use the results after accounting for redundancy. Thus, using the KB, after accounting for redundancy we see a 170% gain when assessing the Internet Census 2012, and a 380% gain for Encore. Note that these gains are in *the quantity of exclusive, novel* information compared to the control approach with **E**. The exact new information provided differs between each approach.

A more accurate comparison between the two methods may be to use the total number of $\in \mathbb{S}$ observations found, including observations labeled as $\emptyset$ (i.e. observations found by both approaches). These gains are more conservative, at +90% for both the Census 2012 and Encore papers. Since the KB approach **T** yields nearly twice as much information as the traditional expert approach **E**, this can be viewed as evidence for the effectiveness of a knowledge base.

In addition, the efficiency (% NEW under *No Redundancy*) was 230% and 1100% greater for the KB tool (**T**) compared to the expert critiques (**E**), for the Internet Census and Encore respectively. This may be the result of **C** including a number of points that can not be said to be directly related to ethics, but were nevertheless purposely included as observations for the sake of making them more coherent as publications. For **E**, in both cases %NEW dropped significantly when adjusting for noise and redundancy, from 46% to 29% for the Census, and from 25% to only 6% in the case of Encore, whereas **T** showed little change in %NEW, remaining high at 97% to 96% with the Census and increasing from 74% to 76% in the case of Encore.

### 6.2 Contributions from the KB

$D^T$ contributed a number of novel observations to the ethical analyses of **O**. We list the major (i.e., $\nvdash$) ones below.

The KB has rules for data collection about computer systems, separate from its human subjects rules. This makes it immune to questions about whether human subjects are involved that could result in a committee stopping short of a thorough ethical analysis. In general the KB goes into more depth on issues and the conditions surrounding them. It also was better at pointing out *Permitted* actions in **O**.

#### 6.2.1 Internet Census 2012

- Collecting MAC addresses of devices is a *Gray* action.
- It is *Recommended* to ensure that traffic your assets send during your research is not malicious. The Carna botnet's traffic can be considered malicious, but it is unclear to what extent its creator assessed this.
- According to Zmap's principles, which have seen significant adoption [28,29,47], it was *Demanded* that Carna's creator coordinate closely with local network admins to reduce risks and handle inquiries (but he did not).
- *Demanded*: Feasibly minimize data collected/stored
- Signing research agreements limiting the use of data to the present experiment is *Recommended*
- Although it is common practice to anonymize data collected from or that could potentially be linked back to *human subjects*, the act of pseudonymizing, anonymizing, or aggregating collected data about *computer systems* being a *Demanded* practice is unique to the KB.
- Encrypting this data in transit is *Recommended*
- $D^T$ *Recommends* testing before deployment for bugs, consistency, safety, etc., in software, hardware, and services to be used on experiment participants. Even though **O** implied some amount of testing for safety before deploying their technology ("After development of most of the code we began debugging our infrastructure. We used a few thousand devices randomly chosen for this purpose." [17]), **C** did not acknowledge this.
- $D^E$ and $D^T$ both mention that disseminating malware is *Prohibited* (although the KB lists it as *Gray* if consent is obtained). However, **C** and the Menlo Report it was based on do not explicitly mention that infecting computers with malware (i.e. installing and/or executing) is *Prohibited*. This point may have been taken as a given by the authors of [26] and therefore excluded.

#### 6.2.2 Encore

- Requiring experiment participants log into, install things onto, or otherwise employ their personal devices, accounts, or systems is a *Gray* action in the KB.
- "Fairly compensate or pay subjects, including crowd-sourced workers, for their contributions" is *Demanded*.

In **O**, the author did not mention compensating the true targets of the study, the site visitors: "For further incentive, we could institute a reciprocity agreement for webmasters: in exchange for installing our measurement scripts, webmasters could add their own site to Encore's list of targets and receive notification about their site's availability from Encore's client population" [21].

- The *Recommended* practice of testing before deployment also came up in **O** but not **C**. However, in Encore's case, they satisfied this recommendation.
- **C** arguably touched on the fallibility of IRBs, but the KB made explicit that adhering to REB/IRB rules but not to other ethics standards is *Prohibited*.
- Data storage and deletion was not touched on in **C**. $D^T$ included the observation that prioritizing any data-driven aspects of research so as to minimize retention time is *Demanded*.
- In addition, minimizing data retention by deleting data as soon as possible after use (even if it must be collected anew in case of error) is *Recommended*.
- Although a related point regarding deception was mentioned in **C**, $D^T$ added that collecting data as part of a separate service not specifically for collecting that data (e.g. which gives bonus incentives), is *Prohibited* when dealing with human subjects, but *Permitted* if the data is solely about computer systems (although not to the exclusion of obtaining informed consent, etc.).

## 7. Discussion

There were some observations in **C** that **T**'s analysis of **O** did not uncover. We detail them all below for completion.

### 7.1 Internet Census 2012 Critique
#### 7.1.1 ⊯

- **C** very interestingly noted that anonymous publication is unethical because of transparency and accountability [26]. This observation was not turned up by the review of top conference papers used to generate the KB. It is possible that such a "meta-insight" about publication itself can only be discovered from papers specifically about ethics, such as **C**.
- Second, **C** touched on the ethics of releasing *source code*, while $D^T$ noted that distributing *malware* is *Gray*. These were deemed separate issues by **L**, so this observation was labeled unique to $D^E$. Such discrepancies can be resolved by adjusting the wording of the KB or using multiple labelers (see Section 8.3).

#### 7.1.2 $+\alpha$

- *Informed consent for releasing others' data publicly.* Although this was also mentioned in $D^T$ *separate* from consent, the KB contains a number of informed consent provisions. In the future, informed consent can be represented as meta-data for relevant nodes.
- The Census's **C** noted that **O** excluded from their scans "systems that could potentially harm Secondary Stakeholders" as an example of not treating all subjects equitably. Relevant to this, the current KB only has data on equitably (or randomly) assigning *test conditions*.

### 7.2 Encore Critique ($+\alpha$ only)

- **C** mentioned that **O** claimed it was infeasible to accurately measure the potential risk of using Encore, and noted that proceeding with research in this situation is ethically questionable. The KB includes a number of practices for *minimizing* risk, but does not explicitly obligate *assessing* risk.
- The abstract principle of meeting users' expectations. Although present in spirit, this is not explicitly addressed in the concrete practices in the KB.

## 8. Limitations

### 8.1 Overlap Between Coverage and Out-of-Scope

Some items were voted $\notin \mathbb{S}$ by the majority if they were e.g. quotes from the Menlo Report that readers might infer recommendations from – i.e. they weren't truly novel.

The main mapping task also mixed in **L**'s judgment about whether the item was out-of-scope. Because items voted $\in \mathbb{S}$ then took **L**'s label, there were two cases from $D^E$ where **L**'s $\notin \mathbb{S}$ decision was defaulted to.

### 8.2 Potential to Overlook Information

As mentioned in Section 5.1.2, observations extracted from **C** were organized in a hierarchical format similar to the KB's. However, we did not include parent items separately from children. Our labeling scheme therefore cannot account for the case where *both* a parent and all of its children are non-redundant and in-scope; although we do not believe there were any observations in $D^E$ like this.

In addition, for Encore, there were observations unique to **C** that referred to a FAQ on the project's website not mentioned in **O**, which we therefore excluded from the results.

### 8.3 Subjectivity in Labeling

Labeling experiments necessarily involve subjectivity [35, 42]. Due to our team size, using more than a single labeler **L** was infeasible. Ideally two or three labelers would resolve differences through discussion or an objective process [18, 37, 46]. For example, [18] had 54 items coded by 2 coders, [37] had 3 coders for 14 questions from 15 interviews, and [46] had 100 topic model labels and 2 coders.

## 9. Conclusion

We find substantial potential in the use of an ethics knowledge base sourced for research papers to provide the same types of insights as an expert analysis using accepted ethics standards, as well as to yield additional concrete insights that are missed when using abstract, subjective guidelines. In this paper we present preliminary evidence that such a knowledge base enables one to create more comprehensive reports of ethical issues than traditional standards do, with greatly increased efficiency. We outline and implement a process to systematically compare results generated from using the two approaches, and find that the knowledge base from [30, 41] can be two to four times more effective at locating ethical issues. We also identify deficiencies in existing standards, including the extent of data handling recommendations and interactions with computer test subjects.

Future work would involve extending this comparison between ethical analysis approaches from the two case studies we used here to three or four, in order to improve its statistical significance; and conducting a more typical user study to measure how various researchers perform when using existing standards versus the KB's decision tree user interface.

## References

[1] 45 CFR § 46.107 IRB membership. (Jul 2018), https://www.law.cornell.edu/cfr/text/45/46.107, library Catalog: www.law.cornell.edu

[2] ACM Code of Ethics and Professional Conduct (Jun 2018), https://ethics.acm.org/

[3] "A Cyber Security Research Ethics Decision Support User Interface." [Online]. Available: https://www.secom.co.jp/isl/en/research/ethics/

[4] Internet Census 2012 - Thoughts (Mar 2013), https://blog.rapid7.com/2013/03/21/internet-census-2012-thoughts/

[5] EC Council Code Of Ethics (Mar 2016), https://www.eccouncil.org/code-of-ethics/

[6] Ethically Aligned Design, First Edition (Dec 2017), https://ethicsinaction.ieee.org/

[7] Case Studies (Jul 2018), https://ethics.acm.org/integrity-project/case-studies/

[8] Committee on the Use of Humans as Experimental Subjects COUHES Committee Members and Staff (2019), https://couhes.mit.edu/committee-members-and-staff

[9] "Cyber-security Research Ethics Dialog & Strategy Workshop (CREDS 2013)." [Online]. Available: https://www.caida.org/workshops/creds/1305/index.xml

[10] "Cyber-security Research Ethics Dialog & Strategy Workshop (CREDS II - The Sequel)." [Online]. Available: https://www.caida.org/workshops/creds/1405/index.xml

[11] IMPACT Cyber Trust Ethos (Feb 2018), https://www.impactcybertrust.org/ethos

[12] The Ethics in AI Institute (Sep 2019), https://www.schwarzmancentre.ox.ac.uk/Page/ethicsinai

[13] Personal data (2019), https://eugdprcompliant.com/personal-data/

[14] USENIX Security '20 Call for Papers (Oct 2019), https://www.usenix.org/sites/default/files/sec20_cfp_101519.pdf

[15] at 23:14, I.T.i.S.F..M..: Researcher sets up illegal 420,000 node botnet for IPv4 internet map (Mar 2013), https://www.theregister.co.uk/2013/03/19/carna_botnet_ipv4_internet_map/

[16] Akiyama, M.: 研究倫理に関して我々の置かれている状況. In: SCIS 2017. SCIS (2017)

[17] Anonymous: Internet Census 2012: Port scanning /0 using insecure embedded devices: Carna Botnet (Mar 2013), http://census2012.sourceforge.net/paper.html

[18] Association for Computing Machinery: Regulators, Mount Up! Analysis of Privacy Policies for Mobile Money Services (2017), oCLC: 255559492

[19] Bailey, M., Dittrich, D., Kenneally, E.: Applying Ethical Principles to Information and Communication Technology Research: A Companion to the Menlo Report (2013)

[20] Bailey, M., Kenneally, E., Dittrich, D., Maughan, D.: The Menlo Report. SSRN Scholarly Paper ID 2145676, Social Science Research Network, Rochester, NY (Mar 2012), https://papers.ssrn.com/abstract=2145676

[21] Burnett, S., Feamster, N.: Encore: Lightweight measurement of web censorship with cross-origin requests. In: In ACM SIGCOMM (2015)

[22] Carlini Pratyush Mishra, N., Vaidya, T., Zhang, Y., Sherr, M., Shields, C., Wagner, D., Zhou, W.: Hidden Voice Commands (2016)

[23] Chatila, R., Havens, J.C.: The IEEE Global Initiative on Ethics of Autonomous and Intelligent Systems. In: Aldinhas Ferreira, M.I., Silva Sequeira, J., Singh Virk, G., Tokhi, M.O., E. Kadar, E. (eds.) Robotics and Well-Being, vol. 95, pp. 11–16. Springer International Publishing, Cham (2019). DOI: 10.1007/978-3-030-12524-0_2,

[24] Council of European Union: Council regulation (EU) no 679/2016 (2014), https://eur-lex.europa.eu/legal-content/EN/TXT/?uri=CELEX:02016R0679-20160504

[25] Dittrich, D.: The ethics of social honeypots. Research Ethics **11**(4), 192–210 (Dec 2015). DOI: 10.1177/1747016115583380,

[26] Dittrich, D., Carpenter, K., Karir, M.: The Internet Census 2012 Dataset: An Ethical Examination. IEEE Technology and Society Magazine **34**(2), 40–46 (Jun 2015). DOI: 10.1109/MTS.2015.2425592

[27] Durumeric, Z., Wustrow, E., Halderman, J.A.: ZMap: Fast Internet-wide Scanning and its Security Applications p. 15 (2013). DOI: 10.5555/2534766.2534818

[28] Ensafi, R., Winter, P., Mueen, A., Crandall, J.R.: Analyzing the Great Firewall of China Over Space and Time. Proceedings on Privacy Enhancing Technologies **2015**(1), 61–76 (2015). DOI: 10.1515/popets-2015-0005,

[29] Holz, R., Amann, J., Mehani, O., Wachs, M., Ali Kaafar, M.: TLS in the Wild: An Internet-wide Analysis of TLS-based Protocols for Electronic Communication. Internet Society (2016). DOI: 10.14722/ndss.2016.23055,

[30] Inagaki, S., Ramirez, R., Shimaoka, M., Magata, K.: Investigation on Research Ethics and Building a Benchmark. The Institute of Electronics, Information and Communication Engineers, Niigata, Japan (Jan 2018), https://www.iwsec.org/scis/2018/program.html

[31] Jackman, M., Kanerva, L.: Evolving the IRB: Building Robust Review for Industry Research p. 17 (2016)

[32] Jim, T.: There is no standard of ethics in computer security research (Aug 2014), http://trevorjim.com/there-is-no-standard-of-ethics-in-computer-security-research/

[33] Kenneally, E., Fomenkov, M.: Cyber Research Ethics Decision Support (CREDS) Tool. pp. 21–21. ACM Press (2015). DOI: 10.1145/2793013.2793017, http://dl.acm.org/citation.cfm?doid=2793013.2793017

[34] Kleinman, A.: The Most Detailed Map Of The Internet Was Made By Breaking The Law (0400), https://www.huffpost.com/entry/internet-map_n_2926934

[35] Lange, R.T.: Inter-rater Reliability. In: Kreutzer, J.S., DeLuca, J., Caplan, B. (eds.) Encyclopedia of Clinical Neuropsychology, pp. 1348–1348. Springer, New York, NY (2011). DOI: 10.1007/978-0-387-79948-3_1203,

[36] Markham, A., Buchanan, E.: Ethical decision-making and Internet research: Recommendations from the AoIR ethics working committee (Version 2.0). AoIR (2012), http://aoir.org/reports/ethics2.pdf

[37] McGregor, S., Charters, P., Holliday, T., Roesner, F.: Investigating the Computer Security Practices and Needs of Journalists (2015)

[38] McMillan, R.: Is It Wrong to Use Data From the World's First 'Nice' Botnet? Wired (May 2013), https://www.wired.com/2013/05/internet-census/

[39] McNamara, P.: Deontic logic. In: Zalta, E.N. (ed.) The Stanford Encyclopedia of Philosophy. Metaphysics Research Lab, Stanford University, summer 2019 edn. (2019), https://plato.stanford.edu/archives/sum2019/entries/logic-deontic/

[40] Narayanan, A., Zevenbergen, B.: No Encore for Encore? Ethical Questions for Web-Based Censorship Measurement. SSRN Electronic Journal (2015). DOI: 10.2139/ssrn.2665148,

[41] R. Ramirez, S. Inagaki, M. Shimaoka, and K. Magata, "A cybersecurity research ethics decision support UI." USENIX Association, Aug. 2020. [Online]. Available: https://www.usenix.org/conference/soups2020/presentation/ramirez

[42] Smyth, P.: Bounds on the mean classification error rate of multiple experts. Pattern Recognition Letters **17**(12), 1253–1257 (Oct 1996). DOI: 10.1016/0167-8655(96)00105-5,

[43] Snoke, T.: Working with the Internet Census 2012 (Oct 2013), https://insights.sei.cmu.edu/cert/2013/10/working-with-the-internet-census-2012.html

[44] Stocker, C., Horchert, J.: Mapping the Internet: A Hacker's Secret Internet Census. Spiegel Online (Mar 2013), https://www.spiegel.de/international/world/hacker-

```
measures-the-internet-illegally-with-carna-botnet-
a-890413.html
```
[45] Thomas, D.R., Pastrana, S., Hutchings, A., Clayton, R., Beresford, A.R.: Ethical issues in research using datasets of illicit origin. In: Proceedings of the 2017 Internet Measurement Conference on - IMC '17. pp. 445–462. ACM Press, London, United Kingdom (2017). DOI: 10.1145/3131365.3131389,

[46] Weinberg, Z., Sharif, M., Szurdi, J., Christin, N.: Topics of Controversy: An Empirical Analysis of Web Censorship Lists. Proceedings on Privacy Enhancing Technologies **2017**(1), 42–61 (2016). DOI: 10.1515/popets-2017-0004,

[47] Zhang, J., Durumeric, Z., Bailey, M., Liu, M., Karir, M.: On the Mismanagement and Maliciousness of Networks. In: NDSS (2014),